\documentclass[11pt,a4paper]{article}
\usepackage[T1]{fontenc}
\usepackage{lmodern}
\usepackage{amsmath,amssymb,amsthm,mathtools}
\usepackage[margin=25mm]{geometry}
\usepackage{microtype}
\usepackage[round,authoryear]{natbib}
\usepackage{enumitem}
\usepackage{xcolor}
\usepackage[colorlinks=true,linkcolor=blue!45!black,citecolor=blue!45!black,urlcolor=blue!45!black]{hyperref}
\usepackage{fancyhdr}
\setlist[enumerate]{itemsep=2pt,topsep=4pt}
\allowdisplaybreaks[2]
\numberwithin{equation}{section}
\newtheorem{theorem}{Theorem}[section]
\newtheorem{corollary}[theorem]{Corollary}
\newtheorem{lemma}[theorem]{Lemma}
\theoremstyle{remark}
\newtheorem{remark}[theorem]{Remark}
\newcommand{\E}{\mathbb E}
\newcommand{\Q}{\mathbb Q}
\newcommand{\R}{\mathbb R}
\newcommand{\F}{\mathcal F}
\newcommand{\1}{\mathbf 1}
\newcommand{\dd}{\,\mathrm d}

\newcommand{\dW}{\,\mathrm dW}
\newcommand{\BS}{\mathrm{BS}}
\newcommand{\new}{\mathrm{new}}
\newcommand{\old}{\mathrm{old}}

\hypersetup{pdftitle={Power-law implied volatility skew at fixed expiries},pdfauthor={Masaaki Fukasawa}}

\title{\bfseries Power-law implied volatility skew at fixed expiries}
\author{Masaaki Fukasawa\\[3pt]
\small Graduate School of Engineering Science, The University of Osaka, Japan}
\date{}

\begin{document}
\maketitle
\thispagestyle{plain}
\begin{abstract}
We revisit the arbitrage argument in Fukasawa (2021), \emph{Volatility has to be rough}, and examine the admissibility of the strategy that adds delta-hedged option positions as a fixed expiry approaches. Under martingale pricing, a deterministic bound on a weighted H\"older seminorm of spot variance of order above the skew parameter gives a common lower bound on wealth for all finite sums of the hedges and their uniform limit. Integrability of the same quantity already rules out power-law implied volatility skew at that expiry.
We construct a martingale pricing model in which the sum of the terminal gains diverges to positive infinity almost surely and the strategy is inadmissible even before expiry. The model produces power-law skew at prescribed expiries while spot volatility is H\"older continuous of every order below one half. More generally, if the H\"older exponent of spot volatility exceeds the skew parameter, the power law can hold almost surely only on a Lebesgue-null set of deterministic expiries.
\end{abstract}
{\small\noindent\textbf{Keywords:} Stochastic volatility; Implied volatility; Deadline effects; H\"older regularity; Admissibility\par}

\section{Introduction}

Rough volatility models link the regularity of spot volatility to the shape of short-dated option smiles. Empirical studies investigate this regularity using high-frequency returns and realised volatility \citep{GatheralJaissonRosenbaum2018,BennedsenLundePakkanen2022,FukasawaTakabatakeWestphal2022,BolkoEtAl2023,ChongTodorov2025,ChongTodorov2026}. In option pricing, rough volatility models can reproduce a negative implied volatility skew whose magnitude grows as a power of the inverse time to expiry. We consider what this power law implies about the regularity of spot volatility as the observation time approaches a fixed expiry.

Section~3 of \emph{Volatility has to be rough} \citep{Fukasawa2021} studies this implication using a strategy that adds a weighted delta-hedged position at each of a succession of trading dates before expiry. If the skew follows the power law and spot variance has a H\"older exponent above the skew parameter, the sum of the terminal gains diverges to positive infinity almost surely. The paper interpreted this divergence as an arbitrage opportunity but left admissibility unverified. A sufficient condition for the contradiction with martingale pricing is a wealth bound that holds uniformly over all finite sums of the positions.

We give a sufficient condition for this common lower bound. We weight the H\"older seminorm of spot variance by the ratio of the maximum stock price to the minimum spot volatility over the trading horizon. Under martingale pricing, a deterministic bound on this quantity gives the same lower bound for every finite sum of the hedges. Their wealth processes converge uniformly through expiry. Integrability of the weighted seminorm already rules out the power law through the option premiums.

The choice of expiries matters. We prove that, under martingale pricing, if the H\"older exponent of spot volatility exceeds the skew parameter, the power law can hold almost surely only on a Lebesgue-null set of deterministic expiries.

At an individual expiry, pathwise H\"older regularity alone permits the power law. We construct a martingale pricing model in which spot volatility is H\"older continuous of every order below one half and the skew follows the power law at prescribed expiries. In the model, spot volatility responds to return shocks during random intervals that end at these expiries. The market observes the start of each interval and allows for the possibility that another will begin before expiry. In this model the wealth of the summed hedge at its trading dates tends to positive infinity almost surely. Its expectation at every such date is zero, so its expected negative part also tends to infinity. The strategy therefore has no deterministic lower bound on wealth before expiry.

\section{The fixed-expiry hedge and admissibility}\label{sec:relation}

\subsection{The fixed-expiry hedge}\label{subsec:fixed-expiry-hedge}

Interest rates are zero. The market filtration $(\F_t)_{t\ge0}$ satisfies the usual conditions. Fix an expiry $T>0$. Let $S$ be a positive continuous semimartingale with positive spot variance $V$ defined by $\dd[\log S]_u=V_u\dd u$, and set $\sigma_u=\sqrt{V_u}$. At observation time $t<T$ and strike $K>0$, let $P_t(T,K)$ and $C_t(T,K)$ be the put and call quotes. Write $k=\log(K/S_t)$ for log-moneyness and $I(t,T,k)$ for the corresponding Black--Scholes implied volatility. As in Section~3 of \citet{Fukasawa2021}, the hedge uses quotes at strikes chosen at the time of trade.

For a random quantity $R_\theta$ and a deterministic scale $b_\theta>0$, write $R_\theta=O_\omega(b_\theta)$ if, almost surely, there exist a finite constant $C(\omega)$ and a positive threshold $\theta_0(\omega)$ such that $|R_\theta|\le C(\omega)b_\theta$ for all $0<\theta<\theta_0(\omega)$. Write $R_\theta=o_\omega(b_\theta)$ if $R_\theta/b_\theta\to0$ almost surely as $\theta\downarrow0$. For a remainder depending on a strike parameter, uniformity on a given set means applying the same definition to the supremum of its absolute value over that set.

Fix $0<H<1/2$. Write $\theta=T-t$ for time to expiry and $z=k/\sqrt\theta$ for scaled log-moneyness. The $H$-power law means that there exist a level $v_t(T)$ and a slope coefficient $\alpha_t(T)$ such that
\begin{equation}\label{eq:power-law}
 I(t,T,z\sqrt\theta)
 =v_t(T)+\alpha_t(T)z\theta^H+o_\omega(\theta^H),
 \qquad \theta=T-t\downarrow0,
\end{equation}
uniformly for $z$ in each compact set, where, almost surely, both $v_t(T)$ and $-\alpha_t(T)$ stay bounded above and bounded away from zero as $t\uparrow T$. The uniform expansion covers the adapted near-the-money strikes in \citet[Section~3]{Fukasawa2021}.

Fix $c>0$ and an integer $n_0$ large enough that $t_n=T-1/n\ge0$ for $n\ge n_0$. Define the two strikes and the number of calls by
\[
 K_n=S_{t_n}e^{-c/\sqrt n},\qquad
 K'_n=S_{t_n}e^{c/\sqrt n},\qquad
 w_n=K_n/S_{t_n}.
\]
At $t_n$, sell one put at $K_n$ and buy $w_n$ calls at $K'_n$, all expiring at $T$. Hold the net premium in the cash account. From $t_n$ to $T$, hold the opposite of the option portfolio's Black--Scholes delta, using the current spot $S_u$, remaining time $T-u$ and the fixed volatility $\sigma_{t_n}$. Denote the put and call deltas used in this hedge by $\Delta^{\mathrm{put},n}_u$ and $\Delta^{\mathrm{call},n}_u$. Write $x^+=\max(x,0)$. The zero-cost terminal gain is
\begin{align}\label{eq:Pi}
 \Pi^n={}&P_{t_n}(T,K_n)-w_nC_{t_n}(T,K'_n)
 -(K_n-S_T)^++w_n(S_T-K'_n)^+\\
 &+\int_{t_n}^T
 \big(\Delta^{\mathrm{put},n}_u-w_n\Delta^{\mathrm{call},n}_u\big)\dd S_u.\notag
\end{align}
The proposed strategy opens $n^{H-1/2}$ units of this hedge at each $t_n$, for every $n\ge n_0$, and holds them to expiry. The total gain at expiry from the positions opened by $t_N$ is
\begin{equation}\label{eq:weighted-hedge-gain}
 Y_N=\sum_{n=n_0}^N n^{H-1/2}\Pi^n.
\end{equation}

Let $D_n=P_{t_n}(T,K_n)-w_n C_{t_n}(T,K'_n)$ be the initial premium difference. At a common Black--Scholes volatility the two weighted premiums cancel by put--call symmetry. Write $v_n=v_{t_n}(T)$, $\alpha_n=\alpha_{t_n}(T)$, and let $\phi$ be the standard normal density. At either strike, the Black--Scholes vega divided by $S_{t_n}/\sqrt n$ is $\phi(c/v_n)+o_\omega(1)$ for a volatility between $v_n$ and the corresponding implied volatility. Thus \eqref{eq:power-law} gives
\[
 D_n=-2c\alpha_n S_{t_n}\phi(c/v_n)n^{-H-1/2}+o_\omega(n^{-H-1/2}).
\]
The coefficient is eventually bounded away from zero along almost every path. If $V$ is almost surely $H_0$-H\"older with $H<H_0\le1/2$, \citet[Lemma~3.1]{Fukasawa2021} gives $\Pi^n-D_n=O_\omega(n^{-H_0-1/2})$. After weighting, the premium contributions dominate a positive multiple of the harmonic series and the hedge errors are summable. This is the argument of \citet[Theorem~3.1]{Fukasawa2021}, which gives $Y_N\to+\infty$ almost surely.

\subsection{Admissibility of the summed hedge}\label{subsec:wealth-bound}
Section~3 of \citet{Fukasawa2021} sought to rule out the coexistence of the $H$-power law and $H_0$-H\"older spot variance with $H_0>H$. The terminal gains $Y_N$ in \eqref{eq:weighted-hedge-gain} tend to $+\infty$ almost surely under these assumptions. The proposed trade is the sum of all the weighted hedges. The paper left its admissibility unverified.

A self-financing strategy is admissible if its discounted wealth stays above a deterministic lower bound throughout trading, almost surely. Write $X^{(N)}$ for the wealth from trades opened through $t_N$, with $X^{(N)}_0=0$ and $X^{(N)}_T=Y_N$. Write $X_u$ for the wealth of the summed hedge at $u<T$. Only finitely many trades have opened by $u$, so
\[
 X_u=0\quad (u<t_{n_0}),\qquad
 X_u=X_u^{(N)}\quad(t_N\le u<t_{N+1}).
\]
In particular, $X_{t_N}$ is wealth at the trading date $t_N$, whereas $Y_N=X_T^{(N)}$ is measured at expiry. To test the arbitrage argument it suffices to have a finite deterministic $b\ge0$ such that
\begin{equation}\label{eq:common-admissibility}
 \inf_{N\ge n_0}\inf_{0\le u\le T} X^{(N)}_u\ge-b
 \qquad\text{almost surely}.
\end{equation}
Under a common martingale measure $\Q$, \eqref{eq:common-admissibility} would make each $X^{(N)}$ a supermartingale \citep{DelbaenSchachermayer1994}. Since $Y_N+b\ge0$, this would give
\[
 +\infty=\E_{\Q}\!\left[\liminf_{N\to\infty}(Y_N+b)\right]
 \le\liminf_{N\to\infty}\E_{\Q}[Y_N+b]\le b.
\]
Section~\ref{subsec:integrable-hedge} gives a sufficient condition for \eqref{eq:common-admissibility}. The model in Section~\ref{sec:model} shows that pathwise H\"older regularity alone does not ensure it.

\subsection{H\"older bounds on spot variance and hedge wealth}\label{subsec:integrable-hedge}
For a path $h$ on $[0,T]$ and exponent $\eta>0$, write $[h]_{\eta;[0,T]}=\sup_{0\le u<v\le T}|h(v)-h(u)|/(v-u)^\eta$. For $H<H_0\le1/2$, define
\[
 G=[V]_{H_0;[0,T]}\frac{\sup_{0\le u\le T}S_u}{\sqrt{\inf_{0\le u\le T}V_u}}.
\]
For options priced by conditional expectation, we take right-continuous price processes at rational strikes and extend them to all strikes by the $1$-Lipschitz bound in $K$. We use these versions throughout.

\begin{theorem}\label{thm:hedge-bounds}
Suppose $S$ is a positive continuous true martingale under $\Q$, with strictly positive continuous spot variance $V$. Suppose the options are priced under $\Q$ by conditional expectation of their payoffs. For the strategy defined in Section~\ref{subsec:fixed-expiry-hedge}, the following hold.
\begin{enumerate}[label=(\roman*)]
\item If $\E_{\Q}G<\infty$, the $H$-power law \eqref{eq:power-law} cannot hold almost surely.
\item If $G\le g$ almost surely for a finite deterministic $g$, the finite wealth sums $X^{(N)}$ satisfy the common lower bound \eqref{eq:common-admissibility}. They converge almost surely, uniformly on $[0,T]$, to a bounded martingale $X$ with the same lower bound. Its terminal value is $X_T=\lim_{N\to\infty}Y_N$ almost surely.
\end{enumerate}
\end{theorem}

\begin{proof}
Assume $\E_{\Q}G<\infty$. Fix $t<T$ and an integrable, positive $\F_t$-measurable strike $K$. Write $P_{\BS}(s,K,\theta,v)$ for the Black--Scholes put price at spot $s$, strike $K$, time to expiry $\theta$ and volatility $v$. For $t\le u<T$, set $P_u^{\BS}=P_{\BS}(S_u,K,T-u,\sigma_t)$. Let $\Delta_u$ and $\Gamma_u$ be its spot delta and gamma. Then
\[
 P_u^{\BS}-P_t^{\BS}=\int_t^u\Delta_v\dd S_v
          +\frac12\int_t^u(V_v-V_t)S_v^2\Gamma_v\dd v,
 \qquad P_T^{\BS}=(K-S_T)^+.
\]
For $t\le v<T$, the Black--Scholes gamma bound and the definition of $G$ give
\begin{align*}
 0\le S_v^2\Gamma_v
 &\le \frac{S_v}{\sigma_t\sqrt{2\pi(T-v)}},\\
 |V_v-V_t|\frac{S_v}{\sigma_t}
 &\le [V]_{H_0;[0,T]}(v-t)^{H_0}
 \frac{\sup_{0\le u\le T}S_u}{\sqrt{\inf_{0\le u\le T}V_u}}
 =G(v-t)^{H_0}.
\end{align*}
Put $C=(2\sqrt{2\pi})^{-1}\int_0^1 x^{H_0}(1-x)^{-1/2}\dd x$. With $x=(v-t)/(T-t)$ it follows that
\begin{align*}
 \frac12\int_t^T |V_v-V_t|S_v^2\Gamma_v\dd v
 &\le \frac{G}{2\sqrt{2\pi}}\int_t^T\frac{(v-t)^{H_0}}{\sqrt{T-v}}\dd v\\
 &=CG(T-t)^{H_0+1/2}.
\end{align*}
Since $0\le P_u^{\BS}\le K$, the delta integral is dominated by $K+CG(T-t)^{H_0+1/2}$, so it extends to a uniformly integrable martingale on $[t,T]$. The call delta integral at the same strike differs by $S_u-S_t$ and also has this martingale property.

For the hedge in \eqref{eq:Pi}, the strikes $K_n$ and $K'_n$ are positive and $\F_{t_n}$-measurable. Since $S$ is a true martingale,
\[
 \E_{\Q}[K_n]=e^{-c/\sqrt n}\E_{\Q}[S_0]<\infty,
 \qquad
 \E_{\Q}[K'_n]=e^{c/\sqrt n}\E_{\Q}[S_0]<\infty.
\]
Define the hedging error of the $n$th portfolio by
\[
 R_n:=\Pi^n-D_n,
 \qquad D_n=P_{t_n}(T,K_n)-w_n C_{t_n}(T,K'_n).
\]
Here $\Pi^n$ is the terminal gain in \eqref{eq:Pi} and $D_n$ is the net option premium received at $t_n$.

Write $P_{t_n}^{\BS}(K_n)$ and $C_{t_n}^{\BS}(K'_n)$ for the put and call prices computed at volatility $\sigma_{t_n}$. For either strike $K$, let $\Gamma_v^n(K)$ be the Black--Scholes gamma at $(S_v,K,T-v,\sigma_{t_n})$ and put
\[
 A_n(K)=\frac12\int_{t_n}^T(V_v-V_{t_n})S_v^2\Gamma_v^n(K)\dd v.
\]
The put and call identities at expiry are
\begin{align*}
 (K_n-S_T)^+-\int_{t_n}^T\Delta_v^{\mathrm{put},n}\dd S_v
 &=P_{t_n}^{\BS}(K_n)+A_n(K_n),\\
 (S_T-K'_n)^+-\int_{t_n}^T\Delta_v^{\mathrm{call},n}\dd S_v
 &=C_{t_n}^{\BS}(K'_n)+A_n(K'_n).
\end{align*}
Substituting these identities into \eqref{eq:Pi} gives
\begin{align*}
 R_n
 &=-\left[(K_n-S_T)^+-\int_{t_n}^T\Delta_v^{\mathrm{put},n}\dd S_v\right]
 +w_n\left[(S_T-K'_n)^+-\int_{t_n}^T\Delta_v^{\mathrm{call},n}\dd S_v\right]\\
 &=\underbrace{-P_{t_n}^{\BS}(K_n)+w_n C_{t_n}^{\BS}(K'_n)}_{=0\ \text{by put--call symmetry}}
 -A_n(K_n)+w_n A_n(K'_n).
\end{align*}
The gamma estimate above is uniform in $K$. Thus, for $K\in\{K_n,K'_n\}$,
\[
 |A_n(K)|\le\frac12\int_{t_n}^T |V_v-V_{t_n}|S_v^2\Gamma_v^n(K)\dd v
 \le CG(T-t_n)^{H_0+1/2}.
\]
Since $0<w_n=e^{-c/\sqrt n}\le1$ and $T-t_n=1/n$,
\[
 |R_n|\le |A_n(K_n)|+w_n|A_n(K'_n)|
 \le (1+w_n)CG(T-t_n)^{H_0+1/2}
 \le 2CGn^{-H_0-1/2}.
\]
To bound $D_n$, take the conditional expectation of \eqref{eq:Pi} given $\F_{t_n}$. The delta integrals have zero conditional mean, and pricing the two payoffs under $\Q$ gives
\begin{align*}
 \E_{\Q}[\Pi^n\mid\F_{t_n}]
 &=D_n-\E_{\Q}[(K_n-S_T)^+-w_n(S_T-K'_n)^+\mid\F_{t_n}]\\
 &=D_n-P_{t_n}(T,K_n)+w_n C_{t_n}(T,K'_n)=0.
\end{align*}
Since $\Pi^n=D_n+R_n$ and $D_n$ is $\F_{t_n}$-measurable,
\begin{equation}\label{eq:integrable-put-error}
 D_n=-\E_{\Q}[R_n\mid\F_{t_n}],\qquad
 |D_n|\le\E_{\Q}[|R_n|\mid\F_{t_n}]
 \le 2Cn^{-H_0-1/2}\E_{\Q}[G\mid\F_{t_n}].
\end{equation}

\emph{(i)} Since $\E_{\Q}[G\mid\F_{t_n}]$ converges almost surely to a finite limit and $H_0>H$, \eqref{eq:integrable-put-error} gives $D_n=o_\omega(n^{-H-1/2})$.
The premium asymptotic from \eqref{eq:power-law} instead gives $\liminf_n n^{H+1/2}D_n>0$ almost surely.

\emph{(ii)} Suppose $G\le g$. Let $X_u^n$ be the wealth of the $n$th hedge, with its option positions valued at their market prices. It is zero before $t_n$. The delta integrals and the option price processes are martingales on $[t_n,T]$, so
\[
 X_u^n=\E_{\Q}[\Pi^n\mid\F_u],\qquad 0\le u\le T,
\]
where the identity before $t_n$ follows from $\E_{\Q}[\Pi^n\mid\F_{t_n}]=0$. The bounds on $R_n$ and $D_n$ give $|\Pi^n|\le |R_n|+|D_n|\le4Cg n^{-H_0-1/2}$. Consequently the wealth of the portfolio with terminal gain $Y_N$ satisfies
\[
 X_u^{(N)}=\sum_{n=n_0}^N n^{H-1/2}X_u^n
 \ge -4Cg\sum_{n=n_0}^{\infty}n^{-1-(H_0-H)}.
\]
For the right-continuous wealth processes the same conditional bound gives $\sup_{u\le T}|X_u^n|\le4Cg n^{-H_0-1/2}$ simultaneously for every $n$, almost surely. Hence
\[
 \sum_{n=n_0}^\infty n^{H-1/2}\sup_{u\le T}|X_u^n|
 \le4Cg\sum_{n=n_0}^\infty n^{-1-(H_0-H)}<\infty.
\]
Thus $X_u:=\sum_{n=n_0}^{\infty}n^{H-1/2}X_u^n$ is the uniform limit on $[0,T]$. Before $T$ it equals the wealth of the positions opened by $u$. At expiry $X_T=\lim_{N\to\infty}Y_N$ almost surely. The same deterministic bound holds for every $X^{(N)}$ and for $X$. Since $X_u^{(N)}=\E_{\Q}[Y_N\mid\F_u]$, dominated convergence gives
\[
 X_u=\E_{\Q}[X_T\mid\F_u],\qquad 0\le u\le T.
\]
\end{proof}

The common bound for the finite sums in part~(ii) suffices for the Fatou argument in Section~\ref{subsec:wealth-bound}. The value $X_T$ is defined by uniform convergence of wealth. Interpreting it as a gain from integrating the aggregate holdings in each asset through $T$ would require a separate integrability condition.

\section{An almost-every-maturity restriction}\label{sec:obstruction}

We next consider how often the fixed-expiry power law can hold when spot volatility has only a pathwise H\"older bound. Fix a horizon $L>0$. Let $W$ be a Brownian motion under $\Q$ and let $S$ have dynamics
\begin{equation}\label{eq:stock}
 \frac{\mathrm dS_v}{S_v}=\sigma_v\dW_v,\qquad S_0>0.
\end{equation}
Write $\E=\E_{\Q}$, $\E_t=\E_{\Q}[\,\cdot\mid\F_t]$ and $\Q_t(A)=\Q(A\mid\F_t)$. The option prices are
\begin{equation}\label{eq:model-prices}
 P_t(T,K)=\E_t[(K-S_T)^+],\qquad C_t(T,K)=\E_t[(S_T-K)^+].
\end{equation}

\begin{theorem}\label{thm:ae}
Suppose \eqref{eq:stock} defines a true martingale on $[0,L]$, and $\sigma$ is adapted, continuous, strictly positive and almost surely $\gamma$-H\"older on $[0,L]$, for some deterministic $\gamma>H$, where $0<H<1/2$. Then there is a deterministic set $E\subset(0,L]$ of full Lebesgue measure such that, for every $T\in E$ and every $M<\infty$,
\begin{equation}\label{eq:ae-result}
 \sup_{|z|\le M}
 |I(T-\theta,T,z\sqrt\theta)-\sigma_{T-\theta}|
 =o_\omega(\theta^H)
 \qquad\text{as }\theta\downarrow0.
\end{equation}
\end{theorem}
The proof is in Appendix~\ref{app:maturity-proof}.

\begin{corollary}\label{cor:all-expiries}
Under the assumptions of Theorem~\ref{thm:ae}, the set of deterministic maturities at which \eqref{eq:power-law} holds almost surely is contained in a deterministic Lebesgue-null set. In particular, the law cannot hold almost surely at every maturity in an interval of positive length.
\end{corollary}

\section{Volatility pulses tied to deadlines}\label{sec:model}

Theorem~\ref{thm:ae} leaves open the possibility of exceptional expiries. We construct a martingale pricing model with such an expiry and show that pathwise H\"older regularity alone does not guarantee admissibility of the summed hedge in Section~\ref{subsec:fixed-expiry-hedge}. The skew follows the $H$-power law at the prescribed expiry even though spot volatility is H\"older continuous of every order below one half. Along the trading dates approaching this expiry, the wealth $X_{t_N}$ tends to $+\infty$ almost surely, while $\E X_{t_N}=0$ for every $N$. The construction also allows a finite or countable set of prescribed expiries.

Fix a horizon $L>0$ and choose
\begin{equation}\label{eq:pq}
 p>0,\qquad q>0,\qquad H=p+q<\tfrac12.
\end{equation}
Let $W$ be a standard Brownian motion under $\Q$. Let $f:[0,\infty)\to\R$ be Lipschitz continuous and supported on $[0,1]$. In particular $f(1)=0$. For a pulse born at $s$ with deadline $R>s$, set
\begin{equation}\label{eq:pulse}
 \xi^{s,R}_t=(R-s)^{q-1/2}
 f\!\left(\frac{t-s}{R-s}\right)(W_t-W_s),\qquad t\ge s.
\end{equation}
Set $\xi^{s,R}_t=0$ for $t<s$. The pulse starts at zero and vanishes at $R$. For $0<u<1$ and $r=R-s$, Brownian scaling gives
\[
 \bigl(\E_{\Q}|\xi^{s,R}_{s+ur}|^2\bigr)^{1/2}
 =r^q\sqrt{u}\,|f(u)|.
\]

Choose pairwise distinct deterministic deadlines $T_j\in(0,L]$ and weights
\begin{equation}\label{eq:countable}
 \ell_j>0,\qquad \ell_*:=\sum_j\ell_j<\infty.
\end{equation}
The index set may be finite or countably infinite. Let $N^j$ be independent inhomogeneous Poisson processes, independent also of $W$, with deterministic intensities
\begin{equation}\label{eq:intensity}
 \lambda_j(s)=
 \begin{cases}
 \ell_j(T_j-s)^{p-1},&0\le s<T_j,\\
 0,&T_j\le s\le L.
 \end{cases}
\end{equation}
Let $\tau_{j,k}$ be the $k$th jump time of $N^j$, and let $(\F_t)$ be the usual augmentation of the filtration generated by $W$ and all the $N^j$. Choose a response function $a\in C^1(\R;(0,\infty))$ and define
\begin{align}
 U_t&=\sum_j\sum_{k=1}^{N^j_t}
 (T_j-\tau_{j,k})^{q-1/2}
 f\!\left(\frac{t-\tau_{j,k}}{T_j-\tau_{j,k}}\right)
 (W_t-W_{\tau_{j,k}}),\label{eq:U}\\
 \sigma_t&=a(U_t).\label{eq:a}
\end{align}
Assume that the stock $S$ in \eqref{eq:stock} is a true martingale on $[0,L]$ and price options by \eqref{eq:model-prices}. When $a$ is bounded above, the martingale condition holds and $S$ has finite moments of every positive order.

For deadline $T_j$, the expected number of activations is
\[
 \E N^j_{T_j}=\int_0^{T_j}\ell_j(T_j-s)^{p-1}\dd s
 =\frac{\ell_j T_j^p}{p}.
\]
Summing over deadlines gives
\begin{equation}\label{eq:total-intensity}
 \Lambda=\E\sum_j N^j_{T_j}
 =\frac1p\sum_j\ell_j T_j^p
 \le\frac{L^p\ell_*}{p}<\infty.
\end{equation}
Thus only finitely many pulses are realised on $[0,L]$ almost surely, even with countably many deadlines.

For a deterministic option maturity $T$, define
\begin{equation}\label{eq:target-weight}
 \ell(T)=
 \begin{cases}
 \ell_j,&T=T_j\text{ for some }j,\\
 0,&T\notin\{T_j\},
 \end{cases}
 \qquad
 C_f=\int_0^1u f(u)\dd u,\qquad
 c_{H,f}=\frac{C_f}{H+3/2}.
\end{equation}

\begin{theorem}\label{thm:construction}
Under the assumptions above, the following conclusions hold.
\begin{enumerate}[label=(\roman*)]
\item The process $U$ is a continuous semimartingale. The processes $U,\sigma$ and $V=\sigma^2$ are almost surely H\"older continuous on $[0,L]$ of every order $0<\gamma<1/2$. If additionally $a'$ has locally bounded variation, then $\sigma$ and $V$ are continuous semimartingales. Every finite collection of European options, priced by \eqref{eq:model-prices}, admits the common martingale measure $\Q$.
\item For every fixed deterministic $T\in(0,L]$ and $M<\infty$,
\begin{equation}\label{eq:skew-expansion}
 \sup_{|z|\le M}\left|
 I(t,T,z\sqrt\theta)-a(U_t)
 -c_{H,f}\ell(T)\frac{a'(U_t)}{a(U_t)}z\theta^H
 \right|=o_\omega(\theta^H),
 \qquad t=T-\theta\uparrow T.
\end{equation}
\item If
\begin{equation}\label{eq:sign-condition}
 C_f a'(U_{T_j})<0\qquad\Q\text{-a.s. for every }j,
\end{equation}
then \eqref{eq:power-law} holds at all prescribed deadlines simultaneously on one event of probability one.
\end{enumerate}
\end{theorem}

The proof is in Appendix~\ref{app:pulse-proof}.

\paragraph{A single deadline.}
Write $T=T_1$, $\ell=\ell_1$ and $\sigma_*=a(0)$. After the final realised activation, Lipschitz continuity and $f(1)=0$ give
\begin{equation}\label{eq:terminal-flatness}
 \sup_{t\le v\le T}|U_v|
 +\sup_{t\le v\le T}|\sigma_v-\sigma_*|
 =O_\omega(T-t).
\end{equation}
Indeed, each remaining coefficient is bounded by a random constant times $T-v$, and there are finitely many pulses. With $\alpha_*=\ell c_{H,f}a'(0)/a(0)$, we obtain $U_T=0$ and
\begin{equation}\label{eq:single}
 I(T-\theta,T,z\sqrt\theta)
 =\sigma_*+\alpha_*\;z\theta^H+o_\omega(\theta^H).
\end{equation}
Negative skew corresponds to $C_f a'(0)<0$. Given $\F_t$, the probability of at least one more activation before $T$ is $1-\exp(-\ell(T-t)^p/p)$.

For a concrete bounded example, take $f(u)=1-u$ on $[0,1]$ and zero for $u>1$, and set
\begin{equation}\label{eq:tanh}
 a(u)=\sigma_0(1-\varepsilon\tanh u),\qquad
 \sigma_0>0,\quad 0<\varepsilon<1.
\end{equation}
With this response, spot volatility is a semimartingale and the stock is a true martingale with finite moments of every positive order. In the single-deadline case, $\alpha_*=-\varepsilon\ell c_{H,f}<0$. Thus Theorem~\ref{thm:construction} gives the hypotheses of Section~\ref{subsec:fixed-expiry-hedge}, and $Y_N\to+\infty$ almost surely. For the wealth of the summed hedge before expiry, Appendix~\ref{app:pulse-proof} gives
\begin{equation}\label{eq:trading-date-wealth}
 X_{t_N}=\sum_{n=n_0}^N n^{H-1/2}D_n+O_\omega(1)\longrightarrow+\infty
 \qquad\text{almost surely}.
\end{equation}
Each $X^{(N)}$ is a zero-cost martingale by \eqref{eq:model-prices}, square integrability of $S$, and boundedness of the stock holding in each hedge. Thus $\E X_{t_N}=\E X^{(N)}_{t_N}=0$. Writing $x^-=(-x)^+$, we obtain
\[
 \liminf_{N\to\infty}\E X_{t_N}^-
 =\liminf_{N\to\infty}\E X_{t_N}^+
 \ge\E\!\left[\liminf_{N\to\infty}X_{t_N}^+\right]=+\infty.
\]
Hence the summed hedge is inadmissible on $[0,T)$. For each $H<H_0<1/2$, the weighted seminorm $G$ is finite almost surely, but Theorem~\ref{thm:hedge-bounds}(i) implies $\E G=\infty$.

\begin{remark}[Economic interpretation]
One interpretation is that a trader begins rebalancing a hedge before an option expires, temporarily changing how spot volatility responds to stock returns. This interpretation is motivated by work on feedback from dynamic hedging \citep{FreyStremme1997} and on stock price clustering at option expiries \citep{NiPearsonPoteshman2005}. Scheduled earnings announcements offer another example of known dates that affect option prices \citep{DubinskyEtAl2019}. In the single-deadline model, the market observes adjustments as they occur. When the total number is unknown, the probability of a further adjustment in the last $\theta$ units of time is of order $\theta^p$, and a late pulse has typical size of order $\theta^q$. Together these give the scale $\theta^{p+q}$ of the implied-volatility correction. If the total is disclosed, the market can rule out further pulses after the last adjustment, and the leading skew term disappears.
\end{remark}

\begin{remark}[Skew stickiness]
The skew stickiness ratio (SSR) measures the instantaneous response of at-the-money implied volatility to a log stock return in units of the at-the-money skew. Let $A_t(v)$ be the sum of the coefficients of $W_v-W_s$ in \eqref{eq:pulse} over births $s\le t$. Since only finitely many pulses are realised, $\sup_{t\le v\le T}|A_t(v)|$ stays bounded as $t\uparrow T$. It tends to zero in the single-deadline example. With several deadlines, a pulse ending after $T$ may still contribute at $T$. For a smooth bounded response, suppose the Brownian sensitivity of moving at-the-money implied volatility per unit log return is bounded by a constant times $\sup_{t\le v\le T}|A_t(v)|$ and the at-the-money skew has a nonzero leading term of order $(T-t)^{H-1/2}$. The SSR then tends to zero. The derivative asymptotic is an additional condition on the finite-difference expansion \eqref{eq:skew-expansion}. For rough Bergomi, the short-maturity limit with observation time fixed is $H+3/2$ \citep[Theorem~3]{Fukasawa2026SSR}. The same limit holds in Gaussian factor basket models under the assumptions of \citet{FukasawaMaedaOgiwara2026}. Thus models with the same skew exponent can have different short-expiry smile dynamics.
\end{remark}

\appendix
\section{Proof of the restriction across expiries}\label{app:maturity-proof}

In both appendices, $C$ and $c$ denote positive deterministic constants that may change between estimates. Subscripts indicate additional parameter dependence. Write $\1_A$ for the indicator of an event $A$. All expectations are under $\Q$. We use $\E_t[X]=\E[X\mid\F_t]$ and $\Q_t(A)=\Q(A\mid\F_t)$.

Lemma~\ref{lem:stopped-smile} controls implied volatility in the stopped model. Lemma~\ref{lem:late-stop} bounds the conditional probability of stopping near expiry, and Lemma~\ref{lem:stopped-transfer} uses that probability to compare the stopped and original implied volatilities. Throughout, $\theta=T-t$ and $z$ denotes scaled log-moneyness, so the original strike is $S_t e^{z\sqrt\theta}$.

\subsection{Stopping the volatility}
Choose $H<\beta<\min(\gamma,1)$ and define
\[
 A_t=\sup_{0\le u<v\le t}
 \frac{|\sigma_v-\sigma_u|}{|v-u|^\beta},\qquad A_0=0.
\]
It is adapted, nondecreasing and continuous. Indeed, the quotient extends continuously to the diagonal with value zero, since it is bounded by $[\sigma]_{\gamma;[0,L]}|v-u|^{\gamma-\beta}$. Its running supremum is therefore continuous.

For integers $m\ge2$, let $\rho_m$ be the first time in $[0,L]$ at which $A_t\ge m$ or $\sigma_t\notin(1/m,m)$, with $\rho_m=\infty$ if there is no such time. On $\{\rho_m>0\}$ set $\sigma_t^{(m)}=\sigma_{t\wedge\rho_m}$. By continuity, the stopped value lies in $[1/m,m]$, including at the first exit. On $\{\rho_m=0\}$ the initial volatility may already lie outside $[1/m,m]$. On this event set $\sigma_t^{(m)}=1$ for all $t$. Thus $1/m\le\sigma_t^{(m)}\le m$, its $\beta$-H\"older seminorm is at most $m$, and $\sigma_t^{(m)}=\sigma_t$ on $\{t<\rho_m\}$. Almost surely $\rho_m>L$ for all sufficiently large $m$. Let $S^{(m)}$ be driven by the same $W$, with the same initial value, and denote its put prices and implied volatility by $P^{(m)}$ and $I^{(m)}$. The argument $k$ in $I^{(m)}(t,T,k)$ means a strike $S_t^{(m)}e^k$. On $\{t<\rho_m\}$, the two spot prices and the corresponding strikes agree.

\subsection{The stopped smile}
\begin{lemma}[Stopped smile]\label{lem:stopped-smile}
For each $m\ge2$, $I^{(m)}$ takes values in $[1/m,m]$. For each $M<\infty$ and sufficiently small $\theta$,
\begin{equation}\label{eq:frozen-IV}
 \sup_{|z|\le M}|I^{(m)}(t,T,z\sqrt\theta)-\sigma_t^{(m)}|
 \le C_{m,M}\theta^\beta.
\end{equation}
\end{lemma}

\begin{proof}
Let $P_{\BS}(s,K,\theta,v)$ be the Black--Scholes put price and $\Gamma_{\BS}=\partial_s^2 P_{\BS}$. For an integrable $\F_t$-measurable strike $K>0$ and a bounded, $\F_t$-measurable frozen volatility $v$ bounded away from zero, put $\Gamma_u=\Gamma_{\BS}(S_u^{(m)},K,T-u,v)$. It\^o's formula gives
\begin{equation}\label{eq:put-gamma-identity}
\begin{split}
 P_t^{(m)}(T,K)-P_{\BS}(S_t^{(m)},K,\theta,v)
 =\frac12\E\!\left[\int_t^T
 \big((\sigma_u^{(m)})^2-v^2\big)(S_u^{(m)})^2
 \Gamma_u\dd u\ \middle|\ \F_t\right].
\end{split}
\end{equation}
The delta integral has zero $\F_t$-conditional mean because the delta is bounded and $S^{(m)}$ is square-integrable. For $v=\sigma_t^{(m)}$ use
\[
 0\le\Gamma_{\BS}(s,K,T-u,v)\le\frac{1}{sv\sqrt{2\pi(T-u)}},
 \quad |(\sigma_u^{(m)})^2-v^2|\le2m^2(u-t)^\beta,
 \quad \E[S_u^{(m)}\mid\F_t]=S_t^{(m)}.
\]
Substitution gives the uniform bound
\begin{align}\label{eq:frozen-price}
 \frac{|P_t^{(m)}(T,K)-P_{\BS}(S_t^{(m)},K,\theta,\sigma_t^{(m)})|}
 {S_t^{(m)}}
 &\le C_m\int_t^T\frac{(u-t)^\beta}{\sqrt{T-u}}\dd u\\
 &\le C_m\theta^{\beta+1/2}.\notag
\end{align}
Taking $v=1/m$ and $v=m$ in \eqref{eq:put-gamma-identity} gives
\[
 P_{\BS}(S_t^{(m)},K,\theta,1/m)\le P_t^{(m)}(T,K)
 \le P_{\BS}(S_t^{(m)},K,\theta,m).
\]
For $K=S_t^{(m)}e^{z\sqrt\theta}$, vega is at least $c_{m,M}S_t^{(m)}\sqrt\theta$ on $|z|\le M$, $v\in[1/m,m]$ for small $\theta$. Dividing \eqref{eq:frozen-price} by the normalised vega bound proves \eqref{eq:frozen-IV}.
\end{proof}

\subsection{The conditional probability of a late stop}
Put
\[
 p_m(t,T)=\Q(t<\rho_m\le T\mid\F_t).
\]
We use its c\`adl\`ag version
\[
 p_m(t,T)=\E[\1_{\{\rho_m\le T\}}\mid\F_t]-\1_{\{\rho_m\le t\}}.
\]
\begin{lemma}[Late stopping]\label{lem:late-stop}
There is a deterministic set $E\subset(0,L]$ of full Lebesgue measure such that for each $T\in E$ and every $m\ge2$,
\begin{equation}\label{eq:stop-small}
 p_m(T-\theta,T)=o_\omega(\theta^{H+1/2})\qquad(\theta\downarrow0).
\end{equation}
\end{lemma}

\begin{proof}
For fixed $m$, write $O_{m,T}$ for a bound with a deterministic constant that may depend on $m$ and $T$. For Lebesgue-almost every $T$, differentiation of the law of $\rho_m$ on $[0,L]$ gives
\begin{equation}\label{eq:stop-law}
 \Q(T-\delta<\rho_m\le T)=O_{m,T}(\delta).
\end{equation}
Intersect these full-measure sets over $m$ to obtain $E$. Fix $T\in E$, take $H+1/2<r<1$, and set $\delta_n=2^{-n}<T$. For $T-\delta_n\le u\le T$,
\[
 p_m(u,T)\le\E[\1_{\{T-\delta_n<\rho_m\le T\}}\mid\F_u].
\]
Doob's inequality for the martingale on the right gives
\begin{equation}\label{eq:Doob}
 \Q\!\left(\sup_{T-\delta_n\le u\le T}p_m(u,T)>\delta_n^r\right)
 \le \delta_n^{-r}\Q(T-\delta_n<\rho_m\le T)
 =O_{m,T}(\delta_n^{1-r}).
\end{equation}
The bounds in \eqref{eq:Doob} are summable. Almost surely, for all sufficiently large $n$,
\[
 \sup_{T-\delta_n\le u\le T}p_m(u,T)\le\delta_n^r.
\]
For $\delta_{n+1}<\theta\le\delta_n$, this gives $p_m(T-\theta,T)\le2^r\theta^r$ for small $\theta$. Since $r>H+1/2$, \eqref{eq:stop-small} follows simultaneously for all $m$ by a countable intersection.
\end{proof}

\subsection{Comparison with the original smile}
\begin{lemma}[Implied-volatility comparison]\label{lem:stopped-transfer}
Fix $T\in E$, $m\ge2$ and $M<\infty$. On $\{\rho_m>T\}$, for sufficiently small $\theta$,
\begin{equation}\label{eq:stopped-IV-comparison}
 \sup_{|z|\le M}|I(t,T,z\sqrt\theta)-I^{(m)}(t,T,z\sqrt\theta)|
 \le C_{m,M}\frac{p_m(t,T)}{\sqrt\theta}
 =o_\omega(\theta^H).
\end{equation}
\end{lemma}

\begin{proof}
On $\{\rho_m>t\}$, the two stock paths coincide through $T$ unless the stop occurs in $(t,T]$. Since each put payoff lies in $[0,K]$,
\begin{equation}\label{eq:put-stop}
 \1_{\{\rho_m>t\}}
 \frac{|P_t(T,K)-P_t^{(m)}(T,K)|}{S_t}
 \le\frac K{S_t}\E[\1_{\{t<\rho_m\le T\}}\mid\F_t]
 =\frac K{S_t}p_m(t,T).
\end{equation}
On $\{\rho_m>T\}$, take $K=S_t e^{z\sqrt\theta}$ in both models. By \eqref{eq:put-stop} and Lemma~\ref{lem:late-stop}, the price difference divided by $S_t$ is $o_\omega(\theta^{H+1/2})$ uniformly for $|z|\le M$. By Lemma~\ref{lem:stopped-smile}, $I^{(m)}$ lies in $[1/m,m]$. Vega is at least $c_{m,M}S_t\sqrt\theta$ for $|z|\le M$ and $v\in[1/(2m),2m]$. Moving from $I^{(m)}$ to either endpoint of this interval changes the normalised put price by at least $c_{m,M}\sqrt\theta/(2m)$. Hence $I$ lies inside the interval for all sufficiently small $\theta$. The mean value theorem with this vega bound and \eqref{eq:put-stop} gives \eqref{eq:stopped-IV-comparison}.
\end{proof}

\begin{proof}[Proof of Theorem~\ref{thm:ae}]
Fix $T\in E$ and $M<\infty$. On $\{\rho_m>T\}$, $\sigma_t=\sigma_t^{(m)}$. Lemmas~\ref{lem:stopped-smile} and \ref{lem:stopped-transfer} give
\[
 \sup_{|z|\le M}|I(t,T,z\sqrt\theta)-\sigma_t|
 \le C_{m,M}\left(\theta^\beta+\frac{p_m(t,T)}{\sqrt\theta}\right)
 =o_\omega(\theta^H),
\]
by $\beta>H$ and Lemma~\ref{lem:late-stop}. Almost surely, $\rho_m>T$ for some integer $m$. Taking a countable intersection over $m$ and integer $M$ gives \eqref{eq:ae-result}. The put-price versions specified in Section~\ref{subsec:integrable-hedge} and the right-continuous version of $p_m$ extend the price bounds from rational times and strikes to all $t<T$ and $K>0$.
\end{proof}

\section{Proof of the pulse construction}\label{app:pulse-proof}

We first prove the path properties in Theorem~\ref{thm:construction}(i). Lemmas~\ref{lem:pulse-moments}--\ref{lem:skew-projection} then give the skew expansion for bounded responses, and Lemma~\ref{lem:response-localisation} extends it to a general response.

\subsection{Regularity of realised paths}\label{subsec:model-argument}

\begin{proof}[Proof of Theorem~\ref{thm:construction}(i)]

For a pulse born at $s$ with deadline $R$, let $c_{s,R}(v)$ be the coefficient of $W_v-W_s$ in \eqref{eq:pulse}, set to zero before birth and after the deadline. Set $A_v=\sum c_{s,R}(v)$ and $B_v=\sum c_{s,R}(v)W_s$, summing over realised births. Then $U_v=A_vW_v-B_v$. Both are adapted processes of finite variation, since the number of births is finite almost surely and each coefficient is Lipschitz after birth. Hence $U=AW-B$ is a semimartingale. At each birth $s$, the jump of $c_{s,R}(v)(W_v-W_s)$ is zero since $W_s-W_s=0$. Each pulse is therefore continuous and H\"older of every order below $1/2$. The same holds for $\sigma$ and $V$ by local Lipschitz continuity. If $a'$ has locally bounded variation, $a$ and $a^2$ are locally differences of convex functions, so the generalised It\^o formula gives the semimartingale claim \citep{RevuzYor1999}. For fixed $T,K$, the pricing formulas \eqref{eq:model-prices} give the martingale property.

\end{proof}

\subsection{Conditional moments of observed and future pulses}
Fix a deterministic expiry $T$ and put $t=T-\theta$. For $t\le v\le T$, let $U_v^{\old}$ sum the pulses born by $t$ and $U_v^{\new}$ sum those born in $(t,v]$. Thus $U_v=U_v^{\old}+U_v^{\new}$ and $U_t^{\old}=U_t$. Put $D_\theta=\sup_{t\le v\le T}|U_v-U_t|$. Write $N_\theta=\sum_j(N_T^j-N_t^j)$ for the number of births in $(t,T]$, including those with deadlines after $T$, and set
\[
 \mathcal G_{t,T}=\F_t\vee\sigma(N_v^j-N_t^j:\ j\text{ ranges over the deadlines},\ t\le v\le T).
\]

\begin{lemma}[Pulse moments]\label{lem:pulse-moments}
For every integer $m\ge1$ there is a deterministic constant $C_m$ such that
\begin{equation}\label{eq:old-moments}
 \E_t\sup_{t\le v\le T}|U_v^{\old}-U_t|^m
 =O_\omega(\theta^{m/2}),
\end{equation}
\begin{equation}\label{eq:new-moments}
 \E_t\sup_{t\le v\le T}|U_v^{\new}|^m\le C_m\theta^{p+mq}.
\end{equation}
Consequently,
\begin{equation}\label{eq:increment-moments}
 \E_tD_\theta^m
 =O_\omega(\theta^{m/2}+\theta^{p+mq}).
\end{equation}
\end{lemma}

\begin{proof}
\emph{Pulses observed by $t$.}
For a pulse born at $s\le t$ with deadline $R$, its change from $t$ to $v$ is
\[
 c_{s,R}(v)(W_v-W_t)
 +\bigl(c_{s,R}(v)-c_{s,R}(t)\bigr)(W_t-W_s).
\]
On $[s,T]$, let $L_{s,R}$ be a Lipschitz constant for $c_{s,R}$ and let $\|c_{s,R}\|_\infty$ be its supremum norm. For a standard Brownian motion $B$ on $[0,1]$, put $d_m=(\E\sup_{0\le u\le1}|B_u|^m)^{1/m}$. Then
\[
 \left(\E_t\sup_{t\le v\le T}|W_v-W_t|^m\right)^{1/m}
 =d_m\sqrt\theta.
\]
For one observed pulse, the conditional $L^m$ norm of the supremum of its change is bounded by
$\|c_{s,R}\|_\infty d_m\sqrt\theta+L_{s,R}\theta|W_t-W_s|$.
Summing over observed births gives a bound of order $O_\omega(\sqrt\theta)$, since the sum over all realised births is finite and $W$ is bounded on $[0,T]$. This proves \eqref{eq:old-moments}.

\emph{Pulses born after $t$.}
Given $\F_t$, the count $N_\theta$ is Poisson with mean
\begin{equation}\label{eq:count-bound}
 \lambda_\theta
 =\frac1p\sum_j\ell_j\bigl(((T_j-t)^+)^p-((T_j-T)^+)^p\bigr)
 \le\frac{\ell_*}{p}\theta^p.
\end{equation}
In particular, $\E_tN_\theta^m=O(\theta^p)$.

The Brownian increments after $t$ are independent of $\mathcal G_{t,T}$. For a pulse born at $s>t$ with deadline $R$, put $r=R-s$. Its Brownian increment runs for at most $\min(r,\theta)$ before the pulse or the option expires. Boundedness of $f$ and Brownian scaling therefore bound the supremum $L^m$ norm of the pulse by
\[
 \left(\E\!\left[\sup_{t\le v\le T}|\xi^{s,R}_v|^m
             \mid\mathcal G_{t,T}\right]\right)^{1/m}
 \le d_m\|f\|_\infty r^{q-1/2}\sqrt{\min(r,\theta)}
 \le C_m\theta^q.
\]
For $r\le\theta$, the middle bound is proportional to $r^q$. For $r>\theta$, use $q<1/2$. Given $\mathcal G_{t,T}$, summing the bounds over the $N_\theta$ births gives
\[
 \E\!\left[\sup_{t\le v\le T}|U_v^{\new}|^m\mid\mathcal G_{t,T}\right]
 \le C_m N_\theta^m\theta^{mq}.
\]
Taking $\E_t$ gives \eqref{eq:new-moments}. Apply $(x+y)^m\le 2^{m-1}(x^m+y^m)$ to the two suprema in \eqref{eq:old-moments} and \eqref{eq:new-moments} to obtain \eqref{eq:increment-moments}.
\end{proof}

\subsection{Scaled returns under a bounded response}
Until the localisation step, assume $0<\underline a\le a\le\overline a$ and that $a'$ is bounded and uniformly continuous.
Set
\[
 a_t=a(U_t),\qquad b_t=a'(U_t),\qquad
 Z=\frac{W_T-W_t}{\sqrt\theta},\qquad
 J=\int_t^T U_v^{\new}\dW_v.
\]
Given $\F_t$, $Z$ is standard normal and independent of the future arrival record. Set $Y=b_tJ/\sqrt\theta$.

\begin{lemma}[Scaled return]\label{lem:return-expansion}
There is a remainder $\mathcal R_\theta$ such that
\begin{equation}\label{eq:return}
 \frac{S_T/S_t-1}{\sqrt\theta}=a_tZ+Y+\mathcal R_\theta,
\end{equation}
with
\begin{equation}\label{eq:remainder}
 \E_t|\mathcal R_\theta|=o_\omega(\theta^H),
 \qquad \E_tY^2\le C\theta^{p+2q}.
\end{equation}
\end{lemma}

\begin{proof}
Let $\omega_a(\delta)=\sup_{|x-y|\le\delta}|a'(x)-a'(y)|$ be the modulus of continuity of $a'$. For every $\delta>0$,
\begin{equation}\label{eq:C1-remainder}
 \E_t[D_\theta\omega_a(D_\theta)]
 \le \omega_a(\delta)\E_tD_\theta
 +\frac{2\|a'\|_\infty}{\delta}\E_tD_\theta^2.
\end{equation}
Since $H<1/2$ and $p+2q>H$, \eqref{eq:increment-moments} gives
\[
 \limsup_{\theta\downarrow0}\frac{\E_t[D_\theta\omega_a(D_\theta)]}{\theta^H}
 \le \omega_a(\delta)\limsup_{\theta\downarrow0}\frac{\E_tD_\theta}{\theta^H}\xrightarrow[\delta\downarrow0]{}0.
\]

Set $R_v=a(U_v)-a_t-b_t(U_v-U_t)$. Define
\[
 \mathcal R_\theta=\frac1{\sqrt\theta}\int_t^T
 \bigl[a(U_v)(S_v/S_t-1)+b_t(U_v^{\old}-U_t)+R_v\bigr]\dW_v.
\]
This gives \eqref{eq:return}. Apply the conditional It\^o isometry to the first two terms and to $Y$.
\begin{align*}
 \frac1\theta\E_t\int_t^T a(U_v)^2(S_v/S_t-1)^2\dd v
 &\le\frac{\overline a^2}{\theta}\int_t^T(e^{\overline a^2(v-t)}-1)\dd v
 =O(\theta),\\
 \frac{b_t^2}{\theta}\E_t\int_t^T |U_v^{\old}-U_t|^2\dd v
 &=O_\omega(\theta),\qquad
 \E_tY^2=\frac{b_t^2}{\theta}\E_t\int_t^T |U_v^{\new}|^2\dd v
 \le C\theta^{p+2q}.
\end{align*}
For the last term, $|R_v|\le D_\theta\omega_a(D_\theta)$ and
\[
 \E_t\left|\frac1{\sqrt\theta}\int_t^T R_v\dW_v\right|
 \le C\E_t\!\left[\left(\frac1\theta\int_t^T R_v^2\dd v\right)^{1/2}\right]
 \le C\E_t[D_\theta\omega_a(D_\theta)]=o_\omega(\theta^H).
\]
These bounds give \eqref{eq:remainder}.
\end{proof}

\subsection{The call payoff near its exercise boundary}
Set $g_z(x)=(x-z)^+$.

\begin{lemma}[Call price]\label{lem:call-payoff}
Uniformly for $|z|\le M$ for each fixed $M<\infty$,
\begin{equation}\label{eq:payoff-error}
 \frac{C_t(T,S_t e^{z\sqrt\theta})}{S_t\sqrt\theta}
 =\E_t[g_z(a_tZ)]+\E_t[Y\1_{\{a_tZ>z\}}]+o_\omega(\theta^H).
\end{equation}
\end{lemma}

\begin{proof}
\[
 |g_z(x+y)-g_z(x)-y\1_{\{x>z\}}|
 \le |y|\1_{\{|x-z|\le|y|\}}.
\]
Here $Y=0$ on $\{N_\theta=0\}$. For any $r>0$, independence of $N_\theta$ and $Z$, and the bounded normal density, imply
\begin{align*}
 \E_t[|Y|\1_{\{|a_tZ-z|\le|Y|\}}]
 &\le\frac{\E_tY^2}{r}
 +r\Q_t(N_\theta>0,\ |a_tZ-z|\le r)\\
 &\le C\theta^{p+2q}/r
 +r\Q_t(N_\theta>0)\Q_t(|a_tZ-z|\le r)
 \le C(\theta^{p+2q}/r+\theta^p r^2).
\end{align*}
At $r=\theta^{2q/3}$ the bound is $O(\theta^{p+4q/3})=o(\theta^H)$, uniformly in $z$.
Set $X_\theta=(S_T/S_t-1)/\sqrt\theta$ and
$z_\theta=(e^{z\sqrt\theta}-1)/\sqrt\theta$. For $|z|\le M$,
$|z_\theta-z|\le C_M\sqrt\theta$. The payoff is 1-Lipschitz in both arguments, so
\[
 \big|\E_t[(X_\theta-z_\theta)^+]-\E_t[g_z(a_tZ+Y)]\big|
 \le |z_\theta-z|+\E_t|\mathcal R_\theta|=o_\omega(\theta^H).
\]
Together the bounds give \eqref{eq:payoff-error}.
\end{proof}

\subsection{The skew coefficient and its dependence on expiry}
Define $B_{T,R}(\theta)=0$ for $R\le t$. For $R>t$, set
\begin{align}
 B_{T,R}(\theta)&=
 \int_t^{T\wedge R}(R-s)^{H-3/2}
 \int_s^T(v-s)f\!\left(\frac{v-s}{R-s}\right)\dd v\dd s,\label{eq:single-B}\\
 B_T(\theta)&=\sum_j\ell_j B_{T,T_j}(\theta).\label{eq:B}
\end{align}
Set $\Psi_T(\theta)=B_T(\theta)/\theta^{H+3/2}$.

\begin{lemma}[Projection and expiry coefficient]\label{lem:skew-projection}
The integral $J$ satisfies
\begin{equation}\label{eq:projection}
 \E[J\mid\F_t\vee\sigma(Z)]=\frac{Z^2-1}{\theta}B_T(\theta),
\end{equation}
and
\begin{equation}\label{eq:coefficient-limit}
 \Psi_T(\theta)\longrightarrow c_{H,f}\ell(T).
\end{equation}
\end{lemma}

\begin{proof}
For $t<u<v<T$, put
$g(u,v)=\sum_{t<s\le u}c_{s,R}(v)$, summing over births $(s,R)$ in the future arrival record. Given $\mathcal G_{t,T}$, the kernel is deterministic and
\[
 J=\int_t^T\int_t^v g(u,v)\dW_u\dW_v.
\]
For Brownian increments $\Delta W_i,\Delta W_j$ over disjoint subintervals of $[t,T]$ of lengths $h_i,h_j$, Gaussian conditioning gives
\[
 \E[\Delta W_i\Delta W_j\mid\mathcal G_{t,T}\vee\sigma(Z)]
 =\frac{h_ih_j}{\theta}(Z^2-1)\qquad(i\ne j).
\]
Applying this to simple kernels and passing to the limit by the It\^o isometry yields
\[
 \E[J\mid\mathcal G_{t,T}\vee\sigma(Z)]
 =\frac{Z^2-1}{\theta}\int_t^T\int_t^v g(u,v)\dd u\dd v
 =\frac{Z^2-1}{\theta}\sum_{t<s\le T}\int_s^T(v-s)c_{s,R}(v)\dd v.
\]
Averaging over the independent future arrival record gives \eqref{eq:projection}.
The intensity and response coefficients multiply to
$(R-s)^{p-1}(R-s)^{q-1/2}=(R-s)^{H-3/2}$.
Writing $F=\|f\|_\infty$ and $r=R-s$, the absolute value of the inner integral is at most $\tfrac F2\min(r^2,\theta^2)$. Since
$r^{H-3/2}\min(r^2,\theta^2)\le\theta^{H+1/2}$,
\begin{equation}\label{eq:B-bound}
 |B_{T,R}(\theta)|\le\tfrac F2\theta^{H+3/2}
 \quad\text{uniformly in }R.
\end{equation}
At $R=T$, changing variables gives the exact identity
\begin{equation}\label{eq:diagonal-B}
 B_{T,T}(\theta)
 =\frac{\theta^{H+3/2}}{H+3/2}\int_0^1u f(u)\dd u
 =c_{H,f}\theta^{H+3/2}.
\end{equation}
If $R<T$ is fixed, the contribution is eventually zero. If $R>T$ is fixed,
\[
 |B_{T,R}(\theta)|
 \le\frac F6(R-T)^{H-3/2}\theta^3
 =o(\theta^{H+3/2}).
\]
By \eqref{eq:B-bound}, each summand of $\Psi_T(\theta)$ is bounded by $F\ell_j/2$ in absolute value. The individual limits therefore give \eqref{eq:coefficient-limit}.
\end{proof}

\paragraph{Implied volatility for a bounded response.}
Let $\phi$ be the standard normal density. Equation \eqref{eq:projection} gives
\begin{align}\label{eq:linear-price}
 \E_t[Y\1_{\{a_tZ>z\}}]
 &=\frac{b_t B_T(\theta)}{\theta^{3/2}}
     \int_{z/a_t}^{\infty}(u^2-1)\phi(u)\dd u\notag\\
 &=\phi(z/a_t)\frac{b_t}{a_t}z\theta^H\Psi_T(\theta).
\end{align}
Write $c_\theta(z,v)$ for the Black--Scholes call price divided by $S_t\sqrt\theta$ at log-moneyness $z\sqrt\theta$ and volatility $v$. Uniformly on compact $z$-sets and for $v$ in a compact subset of $(0,\infty)$,
\[
 c_\theta(z,v)=\E[g_z(vZ)]+O(\sqrt\theta),\qquad
 \partial_v c_\theta(z,v)=\phi(-z/v+v\sqrt\theta/2).
\]
The first estimate follows from the 1-Lipschitz payoff and the $L^1$ expansion
$\E\bigl|(e^{v\sqrt\theta Z-v^2\theta/2}-1)/\sqrt\theta-vZ\bigr|=O(\sqrt\theta)$. Lemmas~\ref{lem:call-payoff} and \ref{lem:skew-projection} give
\[
 c_\theta(z,I(t,T,z\sqrt\theta))-c_\theta(z,a_t)
 =\phi(z/a_t)\frac{b_t}{a_t}z\theta^H\Psi_T(\theta)+o_\omega(\theta^H).
\]
The derivative is bounded below on $|z|\le M$, $v\in[\underline a/2,2\overline a]$. The price difference therefore first brackets $I$ in that interval, then gives $I-a_t=O_\omega(\theta^H)$. The mean value theorem and \eqref{eq:coefficient-limit} now give \eqref{eq:skew-expansion}.

\subsection{Localisation for a general response}
For each integer $m\ge2$, let $\chi_m$ be a smooth compactly supported cutoff with values in $[0,1]$ and equal to one on $[-m,m]$. Set $a_m=\chi_m a+(1-\chi_m)a(0)$. Let $S^{(m)}$ use the same $U,W$ with volatility $a_m(U)$, and write $I^{(m)}$ for its implied volatility.

\begin{lemma}[Localisation]\label{lem:response-localisation}
For each $M<\infty$, on $\{|U_t|\le m-1\}$,
\begin{equation}\label{eq:a-local-IV}
 \sup_{|z|\le M}|I(t,T,z\sqrt\theta)-I^{(m)}(t,T,z\sqrt\theta)|
 \le C_{m,M}\frac{\Q_t(D_\theta\ge1)}{\sqrt\theta}
 =o_\omega(\theta^H).
\end{equation}
\end{lemma}

\begin{proof}
Compare normalised puts
\[
 \bar P_t(k)=\E_t[(e^k-S_T/S_t)^+],
 \qquad
 \bar P_t^{(m)}(k)=\E_t[(e^k-S_T^{(m)}/S_t^{(m)})^+].
\]
Their future return ratios coincide on
$\{|U_t|\le m-1,\ D_\theta<1\}$.
The absolute difference between the two normalised put payoffs is at most $e^k$ on the event where the return ratios differ. Therefore
\begin{equation}\label{eq:a-localisation}
 \1_{\{|U_t|\le m-1\}}
 |\bar P_t(k)-\bar P_t^{(m)}(k)|
 \le \E[e^k\1_{\{D_\theta\ge1\}}\mid\F_t]
 =e^k\Q_t(D_\theta\ge1).
\end{equation}
Choose an integer $r$ with $r/2>H+1/2$ and $p+rq>H+1/2$.
Equation \eqref{eq:increment-moments} yields
\[
 \Q_t(D_\theta\ge1)
 \le \E_tD_\theta^r=O_\omega(\theta^{r/2}+\theta^{p+rq})
 =o_\omega(\theta^{H+1/2}).
\]
Write $[\underline a_m,\overline a_m]$ for the bounds on $a_m$. The bounded-response expansion gives $I^{(m)}=a_m(U_t)+O_\omega(\theta^H)$ uniformly for $|z|\le M$. The normalised price difference in \eqref{eq:a-localisation} is $o_\omega(\theta^{H+1/2})$ for $|z|\le M$ and any fixed $M<\infty$. For small $\theta$, $I^{(m)}$ thus stays a positive distance from the endpoints of $[\underline a_m/2,2\overline a_m]$. On this interval normalised vega is at least $c_{m,M}\sqrt\theta$, so the price bound places $I$ in the same interval. On $\{|U_t|\le m-1\}$, normalised vega inversion gives \eqref{eq:a-local-IV}.
\end{proof}

\begin{proof}[Completion of the proof of Theorem~\ref{thm:construction}]
Along each path, choose an integer $m>|U_T|+2$. Continuity of $U$ ensures that $a_m,a_m'$ agree with $a,a'$ at $U_t$ for all sufficiently late $t$. Apply the bounded-response expansion to $a_m$ and use Lemma~\ref{lem:response-localisation}. Taking a countable intersection over $m$ gives \eqref{eq:skew-expansion}. For each fixed $T$, take the probability-one event for the estimates at rational observation times, with integer strike bounds $M$. Right continuity of prices in $t$, their Lipschitz bound in strike, and continuity of $S$ and $U$ extend the uniform price estimates to all observation times by approximation from the right. Inversion then gives the stated expansion for implied volatility.

Finally, \eqref{eq:sign-condition} and continuity of $a'/a$ give a finite strictly negative limiting coefficient at each deadline. A countable intersection proves (iii).
\end{proof}

\paragraph{Wealth at the trading dates in the single-deadline example.}
We prove \eqref{eq:trading-date-wealth} for the bounded response \eqref{eq:tanh}. Put $u=t_N$, $\theta=T-u=1/N$, and freeze volatility at $\sigma_u$. With $Z=(W_T-W_u)/\sqrt\theta$, the corresponding Black--Scholes terminal stock price is $\widehat S_T=S_u\exp(\sigma_u\sqrt\theta Z-\sigma_u^2\theta/2)$. In Lemma~\ref{lem:return-expansion} the term $Y$ vanishes unless a pulse is born after $u$. By \eqref{eq:count-bound} and \eqref{eq:remainder},
\[
 \E_u|Y|\le(\E_uY^2)^{1/2}\Q_u(N_\theta>0)^{1/2}
 =O(\theta^{(p+2q)/2+p/2})=O(\theta^H).
\]
The scaled returns of $S$ and $\widehat S$ thus differ in conditional $L^1$ by $O_\omega(\theta^H)$, by \eqref{eq:return} and the Gaussian exponential expansion used after \eqref{eq:linear-price}. Since $S_u$ is bounded along each path,
\[
 \E_u|S_T-\widehat S_T|=O_\omega(\theta^{H+1/2}).
\]
This bounds the difference between the market price and the frozen-volatility price of a put or call at any strike, by the Lipschitz property of the payoff.

Fix $H<H_0<1/2$ and let $X_u^n$ be the wealth of one hedge opened at $t_n\le u$. The It\^o identity in the proof of Theorem~\ref{thm:hedge-bounds}, integrated only to $u$, cancels its stock gain against the change in the option values computed at $\sigma_{t_n}$. The absolute gamma integral for each leg is at most $CGn^{-H_0-1/2}$, with $G<\infty$ almost surely. To value each leg at its market price at $u$, first change the Black--Scholes volatility from $\sigma_{t_n}$ to $\sigma_u$, then use the preceding price bound. Vega is at most $S_u\sqrt{\theta/(2\pi)}$ and $|\sigma_u-\sigma_{t_n}|\le[\sigma]_{H_0;[0,T]}n^{-H_0}$. Since $w_n\le1$, these estimates give, with one finite pathwise constant for all $n_0\le n\le N$ and sufficiently large $N$,
\[
 |X_u^n-D_n|\le C(\omega)\bigl(n^{-H_0-1/2}+N^{-1/2}n^{-H_0}+N^{-H-1/2}\bigr).
\]
Here $N^{-1/2}\le n^{-1/2}$. Consequently,
\[
 \left|X_{t_N}-\sum_{n=n_0}^N n^{H-1/2}D_n\right|
 \le C(\omega)\left(2\sum_{n=n_0}^N n^{-1-(H_0-H)}
       +N^{-H-1/2}\sum_{n=n_0}^N n^{H-1/2}\right)=O_\omega(1).
\]
The premium sum diverges by Section~\ref{subsec:fixed-expiry-hedge}, proving \eqref{eq:trading-date-wealth}.

\section*{Declaration of AI use}
The author used AI assistance from the first draft. ChatGPT (GPT-6) proposed the construction of the volatility model in Section~\ref{sec:model}.  OpenAI Codex (GPT-6) developed and drafted parts of the proofs, checked calculations, revised the exposition, and wrote the Lean formalizations. The author takes responsibility for the manuscript, including its mathematical claims and references.

\end{document}